\documentclass[english]{cccconf}
\usepackage[comma,numbers,square,sort&compress]{natbib}
\usepackage{epstopdf}
\begin{document}

\title{Low-Power Wireless Wearable ECG Monitoring Chestbelt Based on Ferroelectric Microprocessor}

\author{Zhendong Ai\aref{scut},
        Zihan Wang\aref{scut}
        Wei Cui\aref{scut}}

\affiliation[scut]{School of Automation Science and Engineering,
        South China University of Technology, Guangzhou 510641, China
        \email{aucuiwei@scut.edu.cn}}
\maketitle

\begin{abstract}
  Since cadiovascular disease (CVD) posts a heavy threat to people's health, long-term electrocardiogram (ECG) monitoring is of great value for the improvement of treatment.
  To realize remote long-term ECG monitoring, a low-power wireless wearable ECG monitoring device is proposed in this paper. The ECG monitoring device, abbreviated as ECGM, is designed based on ferroelectric microprocessor which provides ultra-low power consumption and contains four parts----MCU, BLE, Sensors and Power. The MCU part means circuit of MSP430FR2433, the core of ECGM. The BLE part is the CC2640R2F module applied for wireless transmission of the collected bio-signal data. And the sensors part includes several sensors like BMD101 used for monitoring bio-signals and motion of the wearer, while the Power part consists of battery circuit, charging circuit and 3.3V/1.8V/4.4V power supply circuit. The ECGM first collects ECG signals from the fabric electrodes adhered to wearers' chest, preprocesses the signals to eliminate the injected noise, and then transmit the output data to wearers' hand-held mobile phones through Bluetooth low energy (BLE). The wearers are enabled to acquire ECGs and other physiological parameters on their phones as well as some corresponding suggestions. The novelty of the system lies in the combination of low-power ECG sensor chip with ferroelectric microprocessor, thus achieving ultra-low power consumption and high signal quality.

\end{abstract}

\keywords{ECG, Wearable device, Telemetry monitoring, Ferroelectric Microprocessor}


\section{Introduction}

Cardiovascular disease (CVD) has posed a threat to human’s health caused about 17.7 million
deaths worldwide in 2015, accounting for 31.5\% of the total 56 million deaths. Besides,
the cardiovascular mortality rates in low-income and middle-income nations is much higher
than that in high-income countries \cite{WHS2017}. Therefore, full-day electrocardiogram (ECG)
signals are badly needed for treatment of CVD patients. However, it is too expensive and
impractical for patients to lie still in the hospital for 24-hour ECG monitoring, thus
making quite important the design of suitable long-time ECG monitoring device.

The holter, which is a type of mobile ECG monitoring device for CVD patients nowadays,
works but not good enough. The CVD patients need to paste the electrodes on their skin
and live with at least three leads, sometimes the amount of leads can be even twelve.
There are many other defects wearing the holter for ECG monitoring. The holter is usually
as large as an old radio and doesn't look easy to moving for wearers which means CVD patients
in this article. By the way, the wet Ag-AgCl electrodes the holter uses are not comfortable
and likely suffer from allergy for wearers.

Is there an alternation which owns smaller volume, less bondage and more soothing?
Wearable ECG monitoring device in truly sense may satisfy the CVD patients' expectation.
Nowadays there are several forms of wearable ECG monitoring device including watch shape,
chestbelt shape, underwear shape and so on. A watch shape device seems not a great idea for
continuously ECG monitoring because ECG measuring needs at least two leads.
The underwear shape may be an ideal choice if the technique of flexible fabric electrode
matures enough. It looks like that the chestbelt is the he most appropriate shape, relatively speaking,
for long-time 2-leads or 3-leads ECG monitoring. And how to design a nice product like that
is still worthy of discussion and research.

Teo et al.\cite{Teo2010} proposed wireless sensor nodes based on the low-power system on chip (SoC) which enables the combination between conventional electrodes and continuous ECG monitoring in real time, thus presenting a solution for wearable wireless sensor nodes properly. With the help of planar fashionable circuit board (P-FCB), Yoo et al.\cite{Yoo2010} embedded the wireless system into breast band and integrated a dry P-FCB electrode and signal acquisition circuit on the basis of body sensor networks (BSN) to monitor ECGs in real time. Fensli et al.\cite{Fensli2010} developed a prototype of wireless cardiac monitoring system which makes into practical the transmission of electrode signal to handheld devices wirelessly. Lee et al.\cite{Lee2015} designed such a bio-signal acquisition and classification system with wireless telemetry for BSN, that composed of three chips including a body-end chip, a receiving-end chip, and a classification chip, can correctly diagnose cardiac diseases based on MIT-BIH arrhythmia database and assist the cardiologists in diagnosing their patients. Deepu et al.\cite{Deepu2016} introduced a design for wearable wireless ECG sensors based on a low power 3-lead ECG-on-chip with integrated real-time QRS detection and lossless data compression technology. Satija et al.\cite{Satija2017} implemented IoT-enabled real-time ECG monitoring framework using ECG sensors, Arduino, android phone, Bluetooth, and cloud server. An innovative light-weight ECG signal quality-aware (SQA) method was applied to classify the obtained signal into acceptable or unacceptable categories.

This paper proposes a new design of wearable ECG monitoring device(ECGM) which features
ultra-low power consumption and high signal quality, on the basis of previous research work.
It is a important part of a complete telemetry healthcare system proposed in our previous work.
And the detailed introduction of the designed ECG monitoring device is placed at the following sections.

Apart from this section (Section 1) which introducts the significance of ECG monitoring,
why design the chestbelt ECG monitoringdevice and some researchs on ECG monitoring,
the following sections expound the design of the ECG monitoring device.
Section 2 states the design scheme of the proposed ECG monitoring device.
Section 3 describes the main hardware design, including the parts selection and the circuit design.
Section 4 expounds main software design during the implement of the device.
Section 5 concludes the paper and points out the future research direction.

\section{Design Scheme}

The significance and reason of designing a wearable ECG monitoring chestbelt are stated in the last section.
As for this section, the overall design scheme is introduced. There are lots of
aspects to take into account for designing a wearable ECG monitoring device and the
key one is the signals' quality.
The collected ECG signals, which output as
electrode voltage generally ranging from 0.1 mV to 2.5 mV \cite{Li2017}, are
easy to be disturbed by noise and how to eliminate the signal noise is the first
problem to design a wearable ECG monitoring device. The technique to filter the
noise in ECG signals can be summed up as two ways: hardware filtering and
software filtering. Compared to hardware filtering, Software filtering shows
several apparent advantages, such as low cost, low circuit complexity, high
accuracy, and high flexibility. There are four main types of noise in ECG
acquisition, namely low-frequency baseline drift, power frequency interference,
muscles movement interference, contact interference \cite{Li2017}, and the
efforts put into finding out excellent algorithm to reduce the damage brought
by these noise sources are particularly important when design a ECG monitoring
device.

Another big problem that can't be ignored is the battery life of the designed ECG
device. With no great technique breakthrough in battery energy storage density,
an ultra-low energy consumption feature is crucial for a qualified wearable ECG
monitoring device because users hope the device keep working for a relatively
long time, e.g. a week. The major source of power consumption in a wearable ECG
monitoring device is the wireless transceiver \cite{Deepu2016}. On the basis of
using ultra-low power components as far as possible, therefore, minimizing the
use of transceiver and compressing data flow can effectively reduce the overall
power consumption of the device.

Based on the above analysis, the ECGM should implement the basic work that collects the wearers' bio-signals and sends
them to the mobile phone or any other available monitor-end device.
At the same time, it is nonnegligible to ensure the accuracy and low-power character of the designed ECGM.
A new type of ECG monitor device with wireless data transmission,
high resolution, low power consumption and small volume is preliminarily designed and tested.

\begin{figure}[!htb]
  \centering
  \includegraphics[width=\hsize]{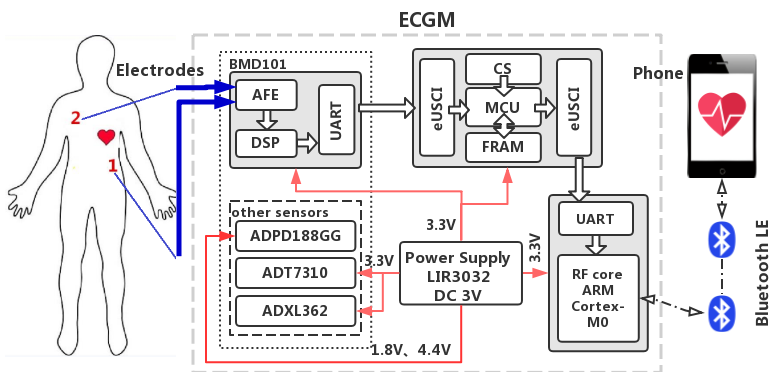}
  \caption{The design scheme of the ECGM}
  \label{ECGM}
\end{figure}

The proposed ECG monitoring device, which named ECGM, contains four main parts, including MCU part for event dealing, BLE part for
wireless transmision, Sensors part for bio-signal acquisition and Power part for different voltage supplys.
Fig.~\ref{ECGM} shows the scheme of the ECGM. The MCU part is actually based on MSP430FR2433,
which is one TI microcontroller with low power and cost, as well as outstanding
performance. The BLE part is designed for wireless communication and the main
device applied is CC2640R2F, one new Bluetooth Low Energy chip fabricated by TI.
The sensor part, which contains ECG sensor and some other sensors which are used
to monitor physiological parameters such as temperature, oxyhemoglobin saturation
and blood pressure. In addition to the mentioned three parts, the power parts
supply several kinds of voltage to the whole device. The working process of the
ECGM can be simply described as: the sensor part collects the wearer's body
information and then send them back to the MCU part. The MCU part send the
wearer's data to the BLE part after the data were received. The BLE convert the data
from UART protocal to BLE protocal and transmit them to the receiver, for example, a
phone with Bluetooth module. The wearer can view their physiological parameters in the
form of number or charts on their mobile phones.

\section{Hardware Design}

The hardware design in this section mainly introduces the four part selection and design of the
proposed ECGM. Limited to limitted length, only important components are expounded and the others
are briefly mentioned.

\subsection{MCU Circuit}

The choice of the MCU is quite significant. There are many factors to consider in
MCU selection: performance, power, stability, cost, compatibility and so on.
Compared to the other serial MCUs like STM32, the MSP430 serial MCU from TI win
a not bad score on the above-mentioned factors.

The MSP430FR2433\cite{CC} microcontroller (MCU) is part of TI’s lowest-cost family of MCUs for sensing and measurement
applications. The architecture, FRAM, and integrated peripherals, combined
with extensive low-power modes, are optimized to achieve extended battery life
in portable and battery-powered sensing applications in a small VQFN package
(4 mm $\times$ 4 mm). What's fantastic is that the MSP430FR2433 combines uniquely
embedded FRAM and a holistic ultra-low-power system architecture, allowing
system designers to increase performance while lowering energy consumption.
FRAM technology combines the low-energy fast writes, flexibility, and endurance
of RAM with the nonvolatility of flash. A significant benefit is that some unchanging
data can be stored in FRAM to lower the power of the MCU while speed up the compute.

\begin{figure}[!htb]
  \centering
  \includegraphics[width=9cm]{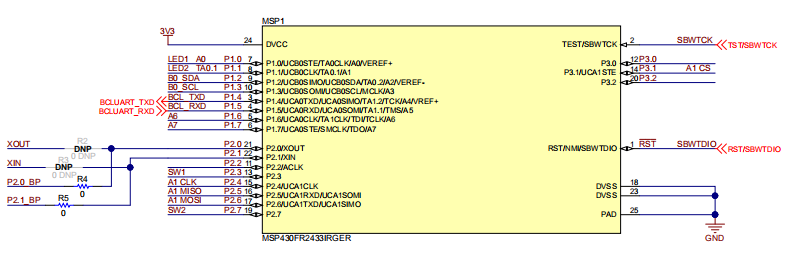}
  \caption{MSP430FR2433 circuit}
  \label{MCU}
\end{figure}

In the proposed ECGM, the MCU acts as a controler and data forwarding. All the used pins
contain power pins, communication pins, LED pins and programming pins. The communication
pins actually means the peripherals used for communication, containning UART and SPI.
UART works for data transmission between BMD101 and MSP430FR2433, while SPI is used for
the others sensors. Actually the two communication ways are implemented in one multiplexed
enhanced universe serial communication interface (eUSCI). One kind of programming
method that MSP430FR2433 support is flash through JTAG interface which is named MSP-FET
in MSP430 serial. Fig.~\ref{MCU} shows the MSP430FR2433 circuit.

\subsection{BLE Circuit}

To transmit the data from MSP430FR2433, a wireless transmission technique is required.
Nowadays there are several popular wireless technique in internet of things
(IoT), namely, WiFi, Bluetooth LE, GPRS, ZigBee, FSK, etc. Considering uses need
ECG monitoring at home instead of hospital, ZigBee network is not a feasible
solution. WiFi brings relatively high power consumption for mobile phone and
ECGM, while GPRS brings additional communication cost. On the whole, Bluetooth
low energy (BLE) seems the best choice for personal ECG monitoring at home.

CC2640R2F\cite{CC} is a bluetooth ultra-low energy wireless MCU with a main CPU, a radio
frequency (RF) core, a sensor controller and other general peripherals/modules.
The chip is designed mainly for Bluetooth 4.2 or Bluetooth 5 low-energy
application. Very low active RF and MCU current and low-power mode current consumption provide excellent battery lifetime and
allow for operation on small coin cell batteries and in energy-harvesting
applications. It's feasible to apply the CC2640R2F as the host MCU and BLE module at the same time,
but in the proposed design, it acts a BLE module only for a better battery life.
In order to smaller the volume of ECGM and speed up product development, a
4mm $\times$ 4mm CC2640R2F BLE module is chose in the design.

\begin{figure}[!htb]
  \centering
  \includegraphics[width=7cm]{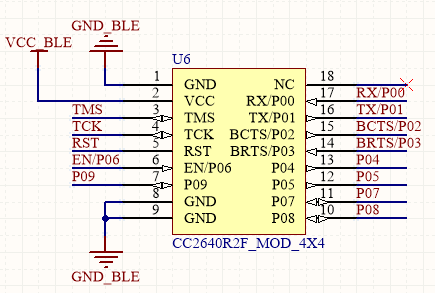}
  \caption{BLE Module Interface}
  \label{BLE}
\end{figure}

The interface circuit of the applied BLE module is showed in Fig.~\ref{BLE}. RX and
TX ports is responsible for data transmission through Bluetooth LE. Except VCC, GND, RX/TX,
there are three important ports which are worthy of attention: EN, BCTS and BRTS.
When the electrical level of EN port goes high, the whole MSP430FR2433 chip powers up.
BCTS port need to set to be high before everytime the BLE module receive some data from
the host MCU and send them out. By the way, once the BLE module transmit data successfully,
the output of BRTS port will be high and become a flag of transfer success for the host MCU.

\subsection{Sensors Circuit}

The sensors parts contains several sensors to monitor wearer's physiological
parameters: BMD101, ADPD188GG, ADT7310 and ADXL362. Actually, because of the
first aim of ECG monitoring, BMD101 is the main sensor introduced in this section.

\subsubsection{BMD101 Circuit}
The BMD101 circuit works as a bio-signal processor in this system. BMD101\cite{BMD},
with advanced analog front-end (AFE) circuitry and a flexible,
powerful digital signal processing (DSP) structure, is
the 3rd generation bio-signal SoC of NeuroSky,

For the AFE part, the main components are a low-noise-amplifier (LNA) and a
16-bit high resolution analog-to-digital converter (ADC). And for the DSP part, there
are a configurable notch filter and a low-pass filter dealing with the ADC data.
Once BMD101 starts working, the collected voltage signal from wearer by the sensor goes
through a high-pass filter and a low-noise-amplifier which are both parts of the AFE.
Then the amplified analog signal is converted to digital signal by the ADC in the AFE part.
After data leaves the ADC, it goes through the digital filters for reducing the negative influence braught
by kinds of noise.
Finally, the row data with
CRC checksum is sent out through universal asynchronous receiver/transmitter
(UART).

Besides, there are several points to notice in the design
of BMD101 circuit. The
first one is the PI filter circuit and it help stable the supply to 3.3V for BMD101.
The two capacitors are both 10$\mu$F while the inductor be selected to be 10$\mu$H.
Another point is that the analog GND and the digital GND should be seperated by an inductor.
In the BMD101 circuit, a 10$\mu$H inductor is employed to reach the above target. In addition to
the above-mentioned tips, there are two TPD1E10B06s, one kind of single channel ESD
protection device in a small 0402 package, are used to protect
the BMD101 chip from damage when the voltage between the two electrodes varies suddenly.

A problem follows the design of BMD101 circuit is to chose right electrodes which are
important for the accurance of the collected ECG signals. For BMD101, there are
some limits in selection of electrodes. Electrodes
suitable for ECG acquisition using BMD101 should be stainless steel,
silver-silver chloride (Ag-AgCl), conductive cloth, etc. To ensure long-time
monitoring and users' comfort, thereby, fabric electrodes are used in this
design. As for dimensions of sensors, an about 10mm-diameter electrode is
recommended.

\subsubsection{Other Sensors}

It is wothy of explanation that there are other sensors applied in the design for
test. Actually a feasible ECG monitoring design should be as lean as possible and
that is why this article titled for ECG monitoring.
The other sensors, including ADPD188GG, ADT7310, ADXL362, is used to acquire other
bio-signals except ECG.

The ADPD188GG\cite{Ana} is a complete photometric system designed to
measure optical signalsfrom ambient light and from synchronous
reflected LED pulses. The module integrates a highly efficient photometric front end, two LEDs, and two photodiode (PD).
The data output and
functional configuration occur over a 1.8 V ${I^2}C$ interface or a
serial peripheral interface (SPI) port.
The ADPD188GG applied in the ECGM is to measure the wearer's heart rate and ${SpO_2}$ through a optical way.

The ADT7310\cite{Ana} is a high accuracy digital temperature sensor
in a narrow SOIC package. It contains a band gap temperature
reference and a 13-bit ADC to monitor and digitize the
temperature to a 0.0625
$^\circ {\rm{C}}$ resolution. The ADT7310 applied in the ECGM is to measure
the wearer's temperature.

The ADXL362\cite{Ana} is an ultra-low power, 3-axis MEMS accelerometer
that consumes less than 2 $\mu$A at a 100 Hz output data rate and
270 nA when in motion triggered wake-up mode. The ADXL362 is used mainly for motion
identification. Once the wearer is in motion, the ECG signal or other bio-signals will
be no longer the same as those acquisited under state of rest. Thus it's necessary
to rectify the deviation for bio-signals under motion state according to the motion data
of the wearer.

\subsection{Power Supply}

At the time the other three parts which mean MCU, BLE and sensors are selected
and designed, the power of the ECGM need to be designed. Cause the 3V battery is
very easy for consumers to reach and the most components used need an input
voltage around 3V, a 3V battery is quitely suitable in the proposed design.

The input voltage required by the components applied in the ECGM are listed in
Table.~\ref{power}.

\begin{table}[!pthb]
  \centering
  \caption{Input Voltage of Components}
  \label{power}
  \begin{tabular}{c|c}
    \hhline
    Component & Input Voltage  \\ \hline

    MSP430FR2433 & 1.8V - 3.6V   \\ \hline

    CC2640R2F & 1.8V - 3.8V   \\ \hline

    BMD101 & 3.3V (+/-10\%)  \\ \hline

    ADPD188GG & 1.7V - 1.9V; 4.0V - 5.0V   \\ \hline

    ADT7310 & 2.7V - 5.5V   \\ \hhline

    ADXL362 & 1.6V - 3.5V   \\ \hhline

  \end{tabular}

\end{table}

It's clear that a 3V DC power is suitable for the ECGM and, to reduce the volume of
ECGM as much as possible, a rechargeable LIR3032 is applied. LIR3032 is one type
of button-cell battery with output voltage range from 2.75V to 4.2V and capacity of
about 120 mAh. In the charging circuit, a MCP73831 chip in charge of the charging of
the LIR3032. The charging current of the LIR3032 should be no more than 0.5CmA and
that means a good charging speed is two hours or longer for a full charge. Thus,
a 16.7 k$\Omega$ or larger resistor is needed to meet the chaging speed limit in the
charging circuit.

Rechargeable feature makes it more convenient for users because they don't have
to buy a new battery from supermarket. What is worth noting is, the voltage will
decrease continuously as the battery is used, proper voltage stabilizing circuit
is necessary. A TPS63036 buck-boost converter capable of providing a
regulated output voltage is applied to generate stable 3.3V output voltage.

Inaddition to the 3.3V voltage, 1.8V and 4.4V is also needed for ADPD188GG.
Taking account of that, a TLV70718 LDO voltage regulator and a
TPS61099 booster converter is used for 1.8V and 4.4V voltage supply, respectively.

\section{Software Design}

The software introduced in this section mainly contains three aspects, the master control program (MCP) for the
data receiving/transmitting through UART/BLE, the SPP program for protocal converting between UART and BLE, as well as ECG data parsing program.
Limited to the article length, other software in the ECGM and the APP for ECG display
are skipped in this section.

\subsection{The master control program}

For the MSP430FR2433, the most important task is receiving the sensing data and then
send them out to the BLE module. The MSP430FR2433 MCU has one active mode and several
software-selectable low-power modes of
operation. An interrupt event can wake the MCU from low-power mode,
service the request, and restore the MCU back to the low-power mode on return from the interrupt
program.

\begin{figure}[!htb]
  \centering
    \includegraphics[width=7cm]{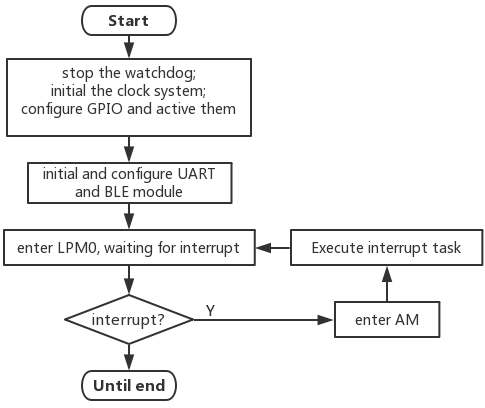}
  \caption{Procedure of MCP}
  \label{MCP}
\end{figure}

The master control program in MSP430FR2433 is designed to wait for the UART interrupt
and then do the interrupt program. A brief procedure of the master control program is showed in Fig.~\ref{MCP} and described
as the following sentences. At the time the MSP430FR2433 powers up, the watchdog timer,
which perform a controlled system restart after a software problem occurs,
will be stopped first. Then the clock system modeule which supports low system cost and low power consumption
should be initialized and the general purpose input and output (GPIO) ports used in the design need to be configured.
After that, it is indispensable to disable the GPIO power-on default high-impedance mode to activate previously configured port settings.
The UART port used need to be initialized and configurated containning selecting the clock source, setting the baudrate
and deciding the UART mode.etc. Of course the BLE module used for data transmission need to be initialized, too.
After all the above steps are performed, the MSP430FR2433 enter hibernation mode (LPM0 in MSP430 serial) and enable
the general interrupt.

When the BMD101 sends out ECG raw data to the UART RX interrupt pin of MSP430FR2433, the host MCU which means MSP430FR2433 awaked, enter the active mode (AM)
and receive the continuous data according to the UART protocal. Samely the host MCU sends the data out through
TX pin to the BLE module after the digital processing of the bio-signal data is done. The procedure described
above is the most directly related to the target of ECG monitoring and some other procedure are skipped here.

\subsection{The SPP Program}

In this design, what the BLE module acts is a RF transceiver. The BLE module receives the data from the host MCU
through UART and then converts them into data obeying the BLE protocal. Of course the BLE data is then transmitted
to the wearer's monitor such as mobile phones.

The application named spp$\_$ble$\_$server in the BLE module is simple and the main processing is
implemented within the application task function.
The application gets the UART data from the SDI layer
and sends it over the air in notification packets.
The application does not directly receive wireless data; the
data goes to the profile layer and gets sent to the UART by the SDI layer. By the way it is feasible to
implement queues to transfer the data to the application layer for further processing.

The initialization of spp$\_$ble$\_$server happens before running the main task and it configures parameters in the
peripheral profile, the GAP, and the GAP bond manager. The initialization
function sets up the serial port service with standard GATT and GAP services in the attribute server and
set the parameters of UART. The registration for receiving UART messagescan be set up
from the SDI layer. Also during this phase, the initialization
function of spp$\_$ble$\_$server calls the
GAPRole$\_$StartDevice function to set up the GAP functions then calls the GAPBondMgr$\_$Register to
register with the bond manager.

\subsection{ECG data parsing}

\begin{figure}[!htb]
  \centering
  \includegraphics[width=7cm]{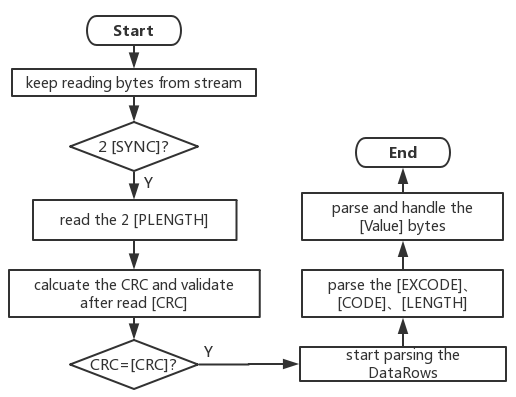}
  \caption{The process of parsing a data packet}
  \label{DATA}
\end{figure}

How to parse the ECG data is important for implement the ECGM and the data parsing
program can run on MSP430FR2433 or on the monitor-end (mobile phones).
BMD101 communicates through UART interface which deploys a 1 start
bit, 8 data bits, and 1 stop bit format. A digital output packet of the UART
interface is sent as an
asynchronous serial stream of bytes. Each packet begins with its Header,
followed by its Data Payload, and ends with its CRC checksum byte. The Data
Payload itself consists of a continuous series of DataRows. Parsing a Data
Payload involves parsing each DataRow until all the bytes of the Data Payload
have been parsed. A DataRow consists of different bytes which are EXCODE, CODE, LENGTH and VALUE respectively.
The process of parsing a ECG digital output packet is illustrated in Fig.~\ref{DATA} and only all the steps described
above are executed correctly in sequence the packet
can be parsed.

\section{Conclusion}

In this paper, a low-power wireless wearable ECG monitoring device in a chestbelt shape is proposed
based on ferroelectric microprocessor MSP430FR2433 and the BMD101 bio-sensor.
The ECG monitoring chestbelt includes four parts, namely, MCU part for event dealing, BLE part for
wireless transmision, Sensors part for bio-signal acquisition and Power part for different voltage supplys.
The main process is: ECG signals are collected, preprocessed, and digitalized by BMD101 sensor, which
acts as a front-end in the ECGM; The digital output
packets are received by MSP430FR2433 through UART, then sent to the BLE module through UART after neccessary
event handling. Then the CC2640R2F converts the UART data to BLE data and transmit them out for the wearer
minitoring their bio-signals especially ECG signal.
Because of the high resolution and low power
consumption of BMD101, ultra-low power consumption and ferroelectric character of MSP430FR2433, the
proposed ECG monitoring device features low power consumption, high signal quality, comfort, etc. By the
way, the main software consist of data parsing, master control program and the SPP program are briefly introduced.
A preliminary design of the proposed ECG monitoring device is complete
and furthermore, more details, such as improving the design and ECG classification algorithm, need to be
studied in the near future.

\end{document}